\documentclass[fleqn,10pt]{wlscirep}
\usepackage[utf8]{inputenc}
\usepackage{float}
\usepackage{bm}
\usepackage{bbold}
\usepackage{amsmath,amssymb}
\usepackage[T1]{fontenc}
\usepackage{url}
\usepackage[ruled,vlined]{algorithm2e}
\title{Bias in Zipf's Law Estimators}

\author[1,*]{Charlie Pilgrim}
\author[2, 3]{Thomas T Hills}
\affil[1]{The University of Warwick, Mathematics for Real-World Systems Centre for Doctoral Training,
Coventry, CV4 7AL, UK}
\affil[2]{The University of Warwick, Department of Psychology, Coventry, CV4 7AL, UK}
\affil[3]{The Alan Turing Institute, British Library, 96 Euston Road, London, NW1 2DB}

\affil[*]{charlie.pilgrim@warwick.ac.uk}

\begin{abstract}
The prevailing maximum likelihood estimators for inferring power law models from rank-frequency data are biased. The source of this bias is an inappropriate likelihood function. The correct likelihood function is derived and shown to be computationally intractable. A more computationally efficient method of approximate Bayesian computation (ABC) is explored. This method is shown to have less bias for data generated from idealised rank-frequency Zipfian distributions. However, the existing estimators and the ABC estimator described here assume that words are drawn from a simple probability distribution, while language is a much more complex process. We show that this false assumption leads to continued biases when applying any of these methods to natural language to estimate Zipf exponents. We recommend that researchers be aware of these biases when investigating power laws in rank-frequency data.
\end{abstract}
\begin{document}

\flushbottom
\maketitle
%
%
\thispagestyle{empty}

\section*{Introduction}
If we take a book and rank each word based on how many times it appears, we will find that the number of occurrences of each word is approximately inversely proportional to its rank \cite{zipf1949human}. The second most frequent word will appear approximately $\frac{1}{2}$ as often as the most frequent word, the third around $\frac{1}{3}$ as frequently. This describes a power law relationship between the frequency of a word, $n$, and the word's rank in terms of its frequency, $r_e$, with exponent $\gamma \approx 1$  \cite{Piantadosi2014Oct},

\begin{equation}
n(r_e) \propto  r_e^{-\gamma} \label{eqn:frequency_dist} \,.
\end{equation}

This is known as Zipf's law and is consistent, in a general sense, across human communication \cite{Ferreri.Cancho2005Mar, Moreno-Sanchez2016Jan}. We do not have a satisfactory reason why this is \cite{Piantadosi2014Oct} and the exponent, $\gamma$, is not always 1 but varies between different speakers \cite{Ferreri.Cancho2005Mar} and texts \cite{Ferreri.Cancho2005Mar, montemurro2002new}. Sound analytical tools are needed to investigate these research areas. 

Equation \ref{eqn:frequency_dist} describes an observed empirical relationship. This is sometimes expressed as a relationship between a word's probability of occurrence \cite{baixeries2013evolution, shannon1951prediction} and the word's rank in the probability distribution, $r_p$, 

\begin{equation}
p(r_p) \propto {r_p}^{-\lambda} \label{eqn:prob_dist} \,.
\end{equation} 

\begin{figure}[ht]
\centering
\includegraphics[width=.6\linewidth]{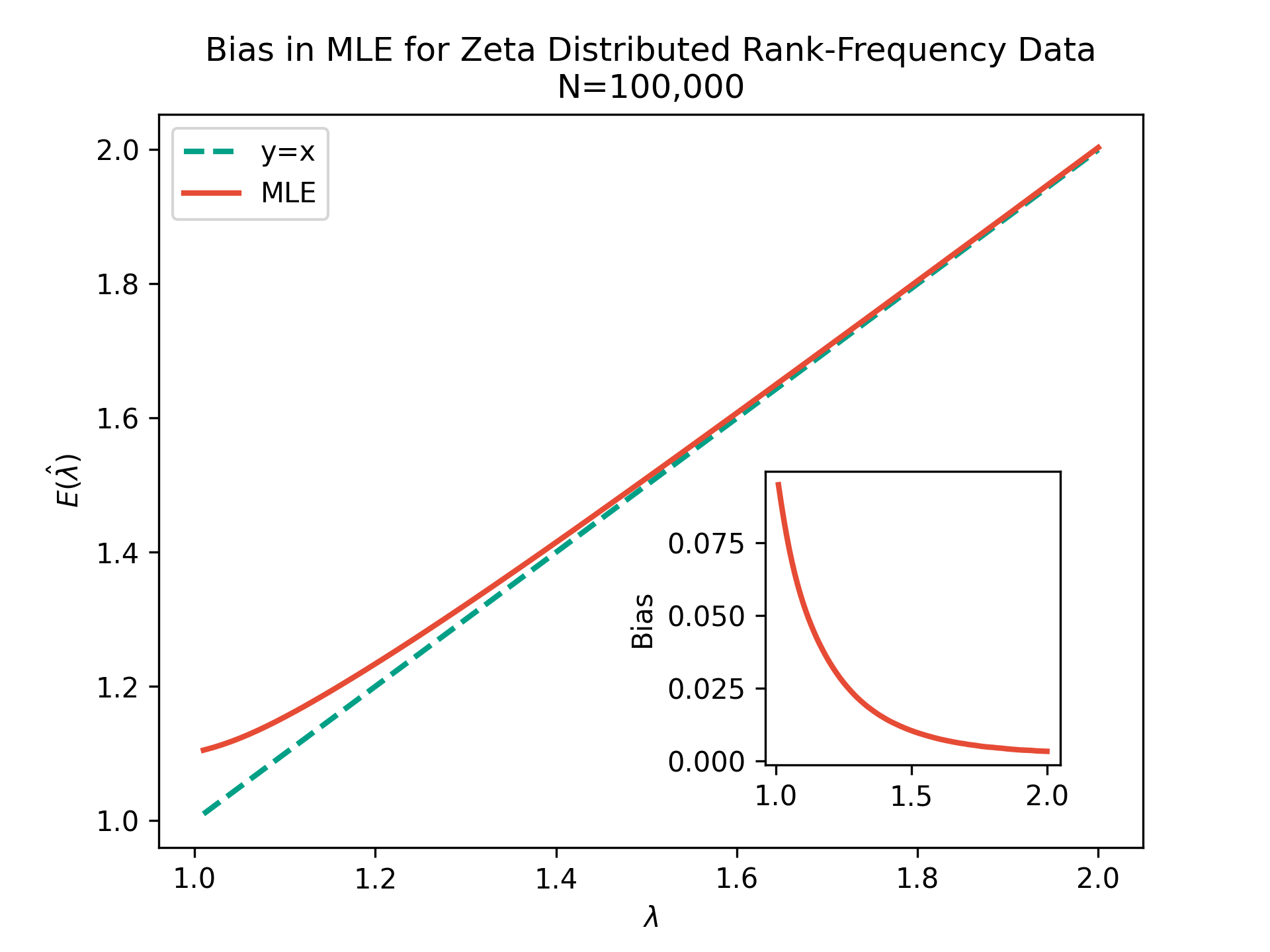}
\caption{Bias in maximum likelihood estimation for rank-frequency data. 100 values of $\lambda$ between 1 and 2 were investigated. For each $\lambda$, samples with $N=100,000$ were generated from an unbounded power law distribution and Clauset et al's estimator was applied to the empirical rank-frequency distribution. This was repeated 100 times and results averaged. There is a clear and strong positive  bias for $\lambda \lessapprox 1.5$.}.  
\label{fig:bias_clauset}
\end{figure}

The conflation of equations \ref{eqn:frequency_dist} and \ref{eqn:prob_dist} causes the prevailing maximum likelihood estimators to miscalculate $\lambda$ in equation \ref{eqn:prob_dist} with a positive bias \cite{Corral2019Aug, Hanel2017Feb} (Figure \ref{fig:bias_clauset}). This bias applies specifically to rank-frequency distributions, where the ranks of events are not known a priori and instead are extracted from the frequency distribution, as is the case in equation \ref{eqn:frequency_dist}. The existing maximum likelihood estimators make the assumption that the observed empirical frequency rankings of data ($r_e$ in equation \ref{eqn:frequency_dist}) are equivalent to rankings in an underlying probability distribution ($r_p$ in equation \ref{eqn:prob_dist}) \cite{Corral2019Aug}, this is the source of the bias.  The $n$th most frequent word is assumed to be the $n$th most likely word, which is not necessarily the case.

\begin{figure}[ht]
\centering
\includegraphics[width=.6\linewidth]{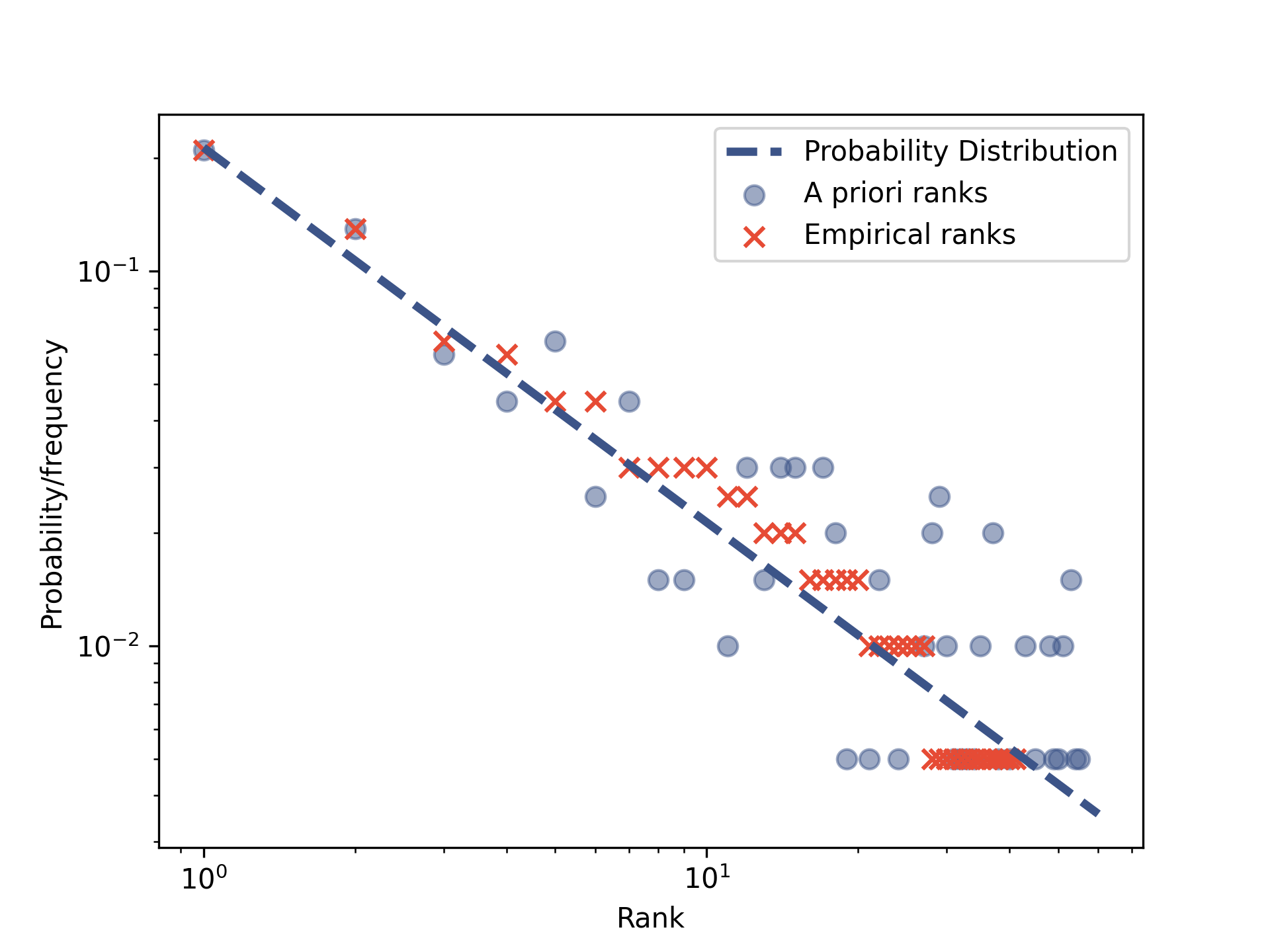}
\caption{ Difference between distributions with probability and empirical ranks. Data was generated from an underlying power law probability distribution with exponent $\lambda=1$, number of possible events $W=60$ and $N=200$ samples. The dotted blue line shows the probability distribution. The blue circles show the sampled event frequencies with a priori known probability ranks. The red crosses show the empirical rank-frequency distribution from the same data. There is a significant difference between the two distributions. The current estimators are designed to fit data with a priori known ranks, not empirical ranks.}.  
\label{fig:rank_and_prob_dist}
\end{figure}

In the 2000s there were a series of papers \cite{clauset2009power, Goldstein2004Feb, Bauke2007Jul, newman2005power}  describing a method of maximum likelihood estimation that gave more accurate (lower bias) estimates for power law exponents than graphical methods \cite{clauset2009power}. The most influential of these is Clauset et al's paper \cite{clauset2009power}. The estimators had been derived and presented before \cite{Goldstein2004Feb} (as early as 1952 in the discrete case \cite{Seal1952}) but Clauset et al's paper popularised the idea and provided a clear methodology including techniques to perform goodness of fit tests \cite{clauset2009power}. In all of these papers, the derivation of the likelihood function assumes that there is some a priori ordering on an independent variable. This works very well for power laws with some natural way to order events, such as the size vs frequency of earthquakes \cite{clauset2009power}. However, it does not work so well with rank-frequency distributions, where the rank is extracted empirically from the frequency distribution, so that the empirical rank and frequency are correlated variables \cite{Piantadosi2014Oct}, both dependent on the same underlying mechanism. This difference was not addressed by Clauset et al, who include examples of applying their estimator to rank-frequency data  \cite{clauset2009power}. The same data can look very different depending on whether we know it's true rank or not, as shown in Figure \ref{fig:rank_and_prob_dist}.

Recently Clauset et al's estimator has been shown, empirically, to be biased for some rank-frequency distributions \cite{Hanel2017Feb, Corral2019Aug}. In particular, Clauset et al's method over-estimates exponents with rank-frequency data generated from known power law probability distributions with exponents below about 1.5 \cite{Hanel2017Feb} (Figure \ref{fig:bias_clauset}). The problem is related to low sampling in the tail \cite{Hanel2017Feb, Corral2019Aug}, so that the observed empirical ranks tend to "bunch up" above the line of the true probability distribution before decaying sharply at the end of the observed tail (Figure \ref{fig:rank_and_prob_dist}). To our knowledge this bias has not been adequately explained or solved.

\begin{itemize}
  \item In 2014 Piantadosi et al \cite{Piantadosi2014Oct} suggested splitting a corpora and calculating ranks of words from one part of the split and frequencies from the other, breaking the correlation of errors. However the method does not take into account uncorrelated errors in the ranks. In particular, the empirical ranks of events in the tail will almost certainly be lower than the actual ranks in the probability distribution as many events in the tail will not be observed at all.
  \item Hanel et al  \cite{Hanel2017Feb} identified the problem and suggested using a finite set of events instead of Clauset et al's unbounded event set \cite{clauset2009power}. This gives more accurate results in the limited case that the number of possible events, $W$, is finite and known \cite{Hanel2017Feb}. Often $W$ is not known and the choice of $W$ can substantially change the results. With Zipf's law in language, $W$ represents the writer's vocabulary and is usually modelled as unbounded \cite{Piantadosi2014Oct, clauset2009power, Bauke2007Jul}. This seems appropriate given that Heaps' Law suggests that the number of unique words in a document continues to rise indefinitely as the document length increases \cite{heaps1978information}.
  \item In 2019 Corral et al \cite{Corral2019Aug} examined the problem and explored a technique of transforming the data to a distribution of frequencies representation, $f(n)$, which is also a power law type distribution that they call the Zipf's law for sizes. This distribution does have an a priori known independent variable of frequency sizes, so the bias described here does not apply to this representation. However there is still difficulty in estimating the rank-frequency exponent, as a power law in the rank-frequency distribution, $n(r_e)$, will only approximately map to a power law in the distribution of frequencies, $f(n)$, for real-world sample sizes \cite{Corral2019Aug}. 
\end{itemize}

Overall these ad-hoc methods can remove the bias to some extent but not completely. The methods also introduce a host of somewhat arbitrary choices for the researcher to resolve. 

We derive a new maximum likelihood estimator that does not make the false assumption that the empirical ranks, $r_e$, are equivalent to the probability ranks, $r_p$. The new estimator considers all the possible ways that the events could be ranked in the underlying probability distribution to generate the observed empirical data. Unfortunately this new likelihood function is computationally intractable for all but the smallest data sets. In order to estimate parameters for larger data sets, we turn to approximate Bayesian computation (ABC), a method that is designed for situations where likelihood functions cannot be computed \cite{Beaumont2010Nov}. We show that this method has much lower bias than Clauset et al's estimator for rank-frequency data generated from simple power laws. We further explore two different implementations of ABC and find that they give different results when applied to word distributions in books because ABC and Clauset et al's method both assume an underlying power law probability model, while natural language arises from a more complex model. We suggest that this false assumption means that maximum likelihood estimation with simple models will always have some arbitrary bias when studying rank-frequency data in natural language, including both ABC and Clauset et al's method. 

\section*{Model}

\subsection*{Likelihood Function - General Case With No A Priori Ordering}

A vector of data, $\bm{d} = [d_1, d_2, ... d_N]$, represents $N$ observations of a random variable $X$. Each of these observations are one of a discrete set of $W$ events, with no a priori ordinality. An example is words in a book. 

We can transform the vector $\bm{d}$ to counts of each event, ordered from most to least frequent, $\bm{n} = [n(x_{(1)}), n(x_{(2)}), . . . , n(x_{(W)})]$. $\bm{n}(x_{(r_e)})$ represents the count of the $r_e$th most common event, where $r_e$ is the event's ranking in the empirical frequency distribution. For ease of notation we will refer to $\bm{n}(x_{(r_e)})$ as $\bm{n}(r_e)$.

We assume a simple model where each of these events has some unknown fixed probability of being observed, $p(x_{r_p}) = Pr(X=x_{r_p})$, where $r_p$ is the event's rank in the underlying probability distribution.

\begin{figure}[H]
\centering
\includegraphics[width=.6\linewidth]{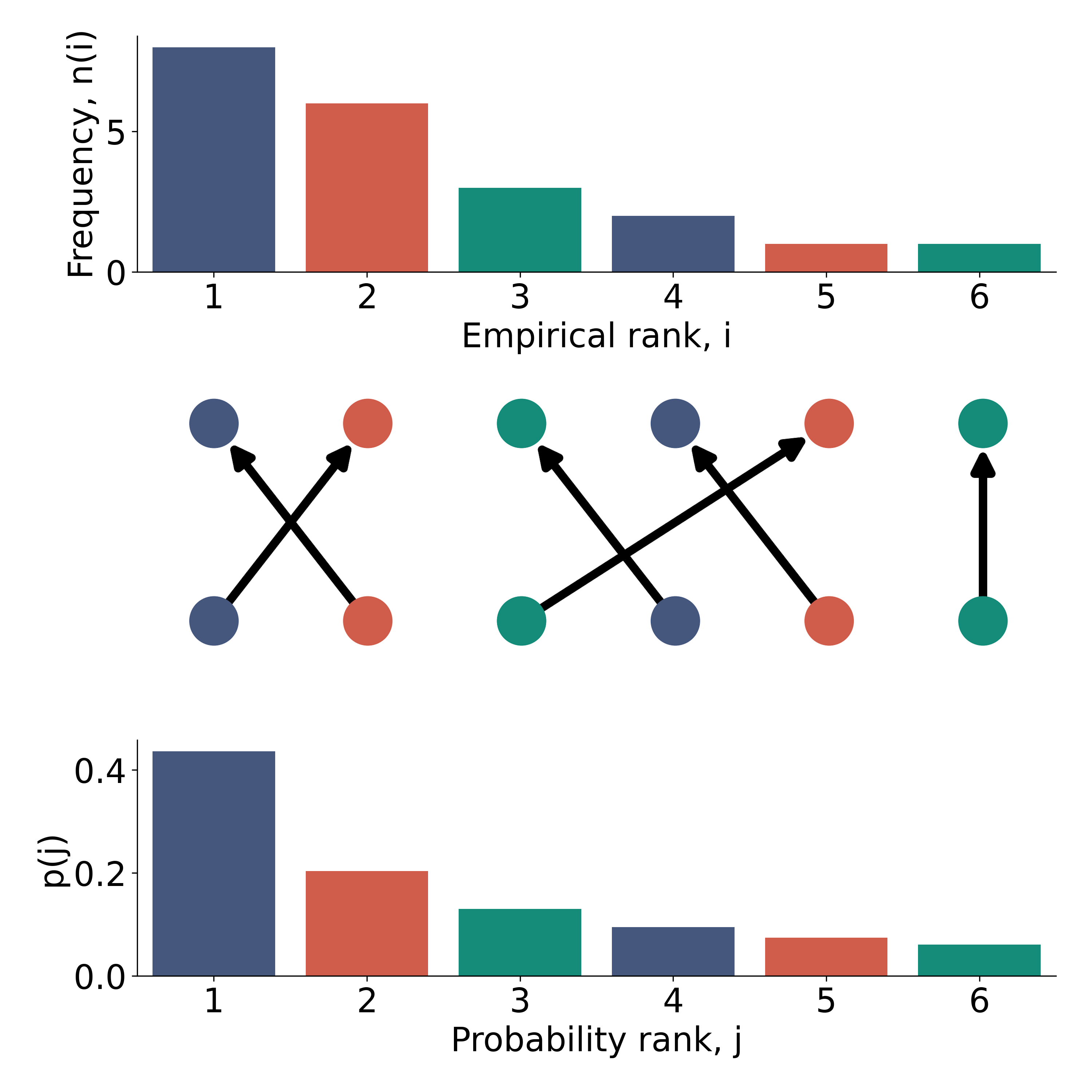}
\caption{An example mapping from probability to empirical ranks. The observed data $\bm{n} = [8,6,3,2,1,1]$ can arise from any valid permutation of events from the probability distribution. Here the permutation is $\bm{s} = [2,1,5,3,4,6]$. The 1st most likely event is observed the second most times ($\bm{s}[1]=2$), etc. The likelihood of the data given this permutation is $p(\bm{n}|\bm{s}, \bm{\theta}) = p_1^{6}p_2^{8}p_3^{1}p_4^{3}p_5^{2}p_6^{1}$}.  
\label{fig:perm_to_prob}
\end{figure}

The key insight is that given an event's empirical rank, we do not know that event’s rank in the underlying probability distribution. We can describe the mapping of events from the data generating probability ranking to the empirical ranking with a vector $\bm{s}$, so that $\bm{s}(r_p) = r_e$. For example $\bm{s} = [2,1,3]$ would mean that the second most probable event was observed empirically the most number of times, the most probable event was seen the second most number of times, and the third most likely seen third most. For any valid mapping, $\bm{s}$ must be a permutation of the integers from 1 to W. Figure \ref{fig:perm_to_prob} shows an example mapping. 

We assume that the probability distribution is parameterised by $\bm{\theta}$. Considering Bayes' rule

\begin{equation}
p(\bm{\theta}|\bm{n}) = \frac{p(\bm{n}| \bm{\theta}) p(\bm{\theta})}{p(\bm{n})} \,.
\end{equation}

The likelihood can be written as (ignoring constants of proportionality)

\begin{equation}
p(\bm{n}| \bm{\theta}) =  \prod_{r_e=1}^W p(x_{(r_e)})^{\bm{n}(r_e)} \,.
\end{equation}

This likelihood equation is in terms of the events' empirical rank, $r_e$, whereas the underlying probability model is in terms of probability rank, $r_p$. To convert the likelihood to be in terms of $r_p$ we condition on the mapping vector, $\bm{s}$,

\begin{equation}
p(\bm{n}|  \bm{\theta}, \bm{s}) = \prod_{r_p=1}^{W} p(x_{r_p})^{\bm{n}(\bm{s}(r_p))} \,.
\end{equation} 

Using the law of total probability we sum over all possible mappings of probability rankings onto empirical rankings. $S(W)$ is the set of all possible permutations of the numbers 1 to W, known as the symmetric group, 

\begin{equation}
p(\bm{n}| \bm{\theta}) =  \sum_{\bm{s} \in S(W)} \prod_{r_p=1}^{W} p(x_{r_p})^{\bm{n}(\bm{s}(r_p))} \label{eqn:likelihood} \,.
\end{equation}

Equation \ref{eqn:likelihood} is the likelihood for any data that represents observations of discrete events, where the events have no a priori ordering in relation to the underlying model. The equation generalises to $W \to \infty$, suitable to describe models with unbounded event sets, as is the case in many Zipf type models. 

\subsection*{Likelihood Function - Power Laws With No A Priori Ordering}

A common model applied to rank-frequency distributions is the power law, used by Zipf in his study of words\cite{zipf1949human}. A power law probability distribution is of the form

\begin{equation}
p(x_{r_p}) = \frac{r_p^{-\lambda}}{Z_{\lambda}} \label{eqn:powerlaw} \,,
\end{equation}

where $\lambda$ is the power law exponent, $Z_{\lambda}$ is a normalising factor. We use the simplest form of Zipf's law for ease of analysis. The method described here can be used with other models such as the Zipf-Mandelbrot law \cite{mandelbrot1953informational}. The normalising factor is

\begin{equation}
    Z_{\lambda} = \sum_{r_p=1}^{W} r_p^{-\lambda} \,,
\end{equation}

where $W$ is the number of possible events. In the limit $W \to \infty$, $Z_{\lambda}$ becomes the Riemann zeta function, $\zeta(\lambda)$ \cite{clauset2009power}.

Considering equation \ref{eqn:likelihood}, the likelihood can be written as

\begin{equation}
\mathcal{L}(\lambda|\bm{n}) =  \sum_{\bm{s} \in S(W)} \prod_{r_p}^{W} \left(\frac{r_p^{-\lambda}}{Z_{\lambda}} \right)^{\bm{n}(\bm{s}(r_p))} \label{eqn:plaw_likelihood} \,.
\end{equation}

And the differential of the likelihood with respect to $\lambda$ is 

\begin{equation}
 \frac{\partial }{\partial \lambda} \mathcal{L}(\lambda|\bm{n}) = \sum_{\bm{s} \in S(W)} \left( \left( \frac{NZ'_{\lambda}}{Z_\lambda} +  \sum_{r_p}^W \bm{n}(\bm{s}(r_p)) ln(r_p)   \right) \times \prod_{r_p}^W \left(\frac{{r_p}^{-\lambda}}{Z_{\lambda}} \right)^{\bm{n}(\bm{s}(r_p))}  \right) \label{eqn:plaw_d_estimator} \,,
\end{equation}

where $Z_{\lambda}'$ is the differential of the normalising factor with respect to $\lambda$. 

\begin{figure}[ht]
\centering
\includegraphics[width=.6\linewidth]{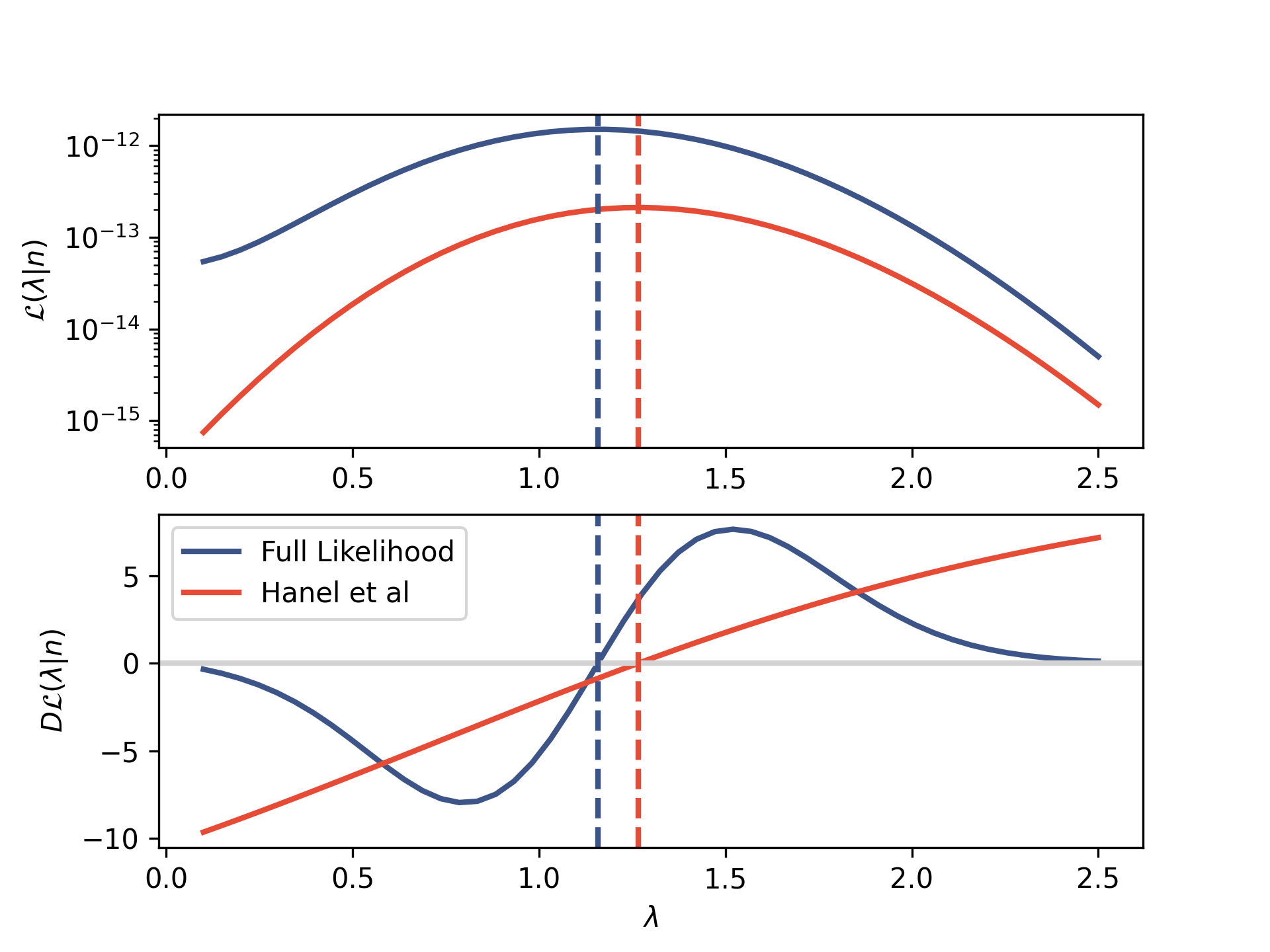}
\caption{Likelihood functions of the full likelihood (blue) and only the leading term (red). Both likelihoods are calculated for the data $\bm{n} = [10,3,3,2,1,1]$. The leading term of the full likelihood is equivalent to the likelihood function as defined by Hanel et al \cite{Hanel2017Feb}, which is adapted for finite event sets from Clauset et al's estimator \cite{clauset2009power}. The top figure shows the full likelihood compared to Hanel et al's likelihood, with the maximum likelihood estimators shown as dashed lines. The bottom figure shows the differential of the likelihood functions. The form of the differential of the full likelihood is markedly different to only the first term. There is a substantial difference in the maximum likelihood estimator, with the Hanel et al estimator giving $\hat{\lambda}= 1.27$ and the full estimator giving $\hat{\lambda}= 1.16$. }
\label{fig:likelihood_function}
\end{figure}

To find the maximum likelihood estimator, we can use numerical methods to either a) maximise equation \ref{eqn:plaw_likelihood} or b) find the root of equation \ref{eqn:plaw_d_estimator} (Figure \ref{fig:likelihood_function}). 

The prevailing estimators from the literature (often implicitly) assume that the empirical ranks match the probability ranks \cite{Piantadosi2014Oct, clauset2009power, Bauke2007Jul}, so that they only consider the leading term in the main sum in both equations \ref{eqn:plaw_likelihood} and \ref{eqn:plaw_d_estimator} (associated with the identity permutation $\bm{s}_I = [1,2,...,W]$). This is the source of the bias in the existing estimators.

The number of terms in the likelihood function (equation \ref{eqn:likelihood}) scales as $O(W!)$, so that naive computation of the likelihood is impractical even at $W \approx 10$. The computation can be shown to be equivalent to the computation of the permanent of a matrix with entries $a_{ij}=p(x_j)^{\bm{n}(i)}$. The best known algorithm for exactly computing the permanent of a matrix is Ryser's algorithm \cite{ryser1963combinatorial, Glynn2010Oct} with complexity $O(W2^W)$. This is computationally intractable for real world data sets such as text corpora with vocabularies of $W>1000$. A more in-depth discussion on the computational complexity can be found in the Supplementary Information.

\subsection*{Approximate Bayesian Computation}

Approximate Bayesian computation is a technique for approximating posterior distributions without calculating a likelihood function \cite{Sunnaker2013Jan, Beaumont2002Dec, Csillery2010Jul}. Instead, we assume a model, $\mathcal{M}$, simulate data, $\bm{n}_i$, from possible parameters, $\lambda_i$, and observe how close that simulated data is to the empirical data using a distance measure $\rho(\bm{n}_i, \bm{n}_{obs})$ \cite{Sunnaker2013Jan, Csillery2010Jul}. The ABC rejection algorithm is based upon the principle that we can approximate the actual posterior by estimating the probability of $\lambda$ given that the data is within some small tolerance, $\epsilon$, of the observed empirical data \cite{Sunnaker2013Jan, Sisson2007Feb}. This assumes that the model, $\mathcal{M}$, is a good representation of the actual data generating process. 

\begin{equation}
    p(\lambda |\bm{n} = \bm{n}_{obs}, \mathcal{M}) \approx p(\lambda|\rho(\bm{n}, \bm{n}_{obs}) < \epsilon, \mathcal{M}) \label{eqn:abc_principle}
\end{equation}

\begin{equation}
    p(\lambda|\rho(\bm{n}, \bm{n}_{obs}) < \epsilon, \mathcal{M}) = \dfrac{p(\rho(\bm{n}, \bm{n}_{obs}) < \epsilon|\lambda, \mathcal{M})p(\lambda | \mathcal{M})}{p(\rho(\bm{n}, \bm{n}_{obs}) < \epsilon | \mathcal{M})} \label{eqn:abc_bayes}
\end{equation}

The ABC rejection algorithm begins by sampling parameter values from the prior. For each of these parameter values, data is then generated from the model and tested on the condition $\rho(\bm{n}_i, \bm{n}_{obs}) < \epsilon$ \cite{Sunnaker2013Jan}. With enough samples, the density of successful parameters will approximate the right hand side of Equation \ref{eqn:abc_bayes}, and an approximation for the posterior distribution \cite{Sunnaker2013Jan}. If we use a uniform prior then this will be a proportional estimate to the likelihood. 

An ideal distance measure, $\rho(\bm{n}_i, \bm{n}_{obs})$, would involve comparing Bayesian sufficient summary statistics from the data \cite{Csillery2010Jul}. Usually in practice Bayesian sufficiency cannot be achieved \cite{Csillery2010Jul, Sunnaker2013Jan}, and some information will be lost so that the approximation of the posterior includes some error\cite{Sunnaker2013Jan}. A common technique is to summarise the data sets with summary statistics, $\bm{S}(\bm{n})$, and define the distance as the difference between those, $\rho(\bm{n}_i, \bm{n}_{obs}) = \bm{S}(\bm{n}_i) - \bm{S}(\bm{n}_{obs})$ \cite{Beaumont2010Nov, Sunnaker2013Jan, Csillery2010Jul}. Recently the Wasserstein distance, a metric between distributions, has been shown to work well as a distance measure \cite{bernton2019approximate}. This is a principled approach that avoids the difficult selection of summary statistics \cite{bernton2019approximate}, and this is the measure that we use here. 

The ABC rejection algorithm requires a small tolerance in order to find a good estimate for the posterior \cite{Sisson2007Feb}. This in turn requires a high density of samples in order to have enough successful parameters to build the posterior approximation. To sample at a high density across a reasonable parameter space with a uniform prior would be prohibitively computationally expensive. Instead, we use population Monte Carlo to sample from a proposal distribution that focuses on areas of high posterior probability while avoiding areas of negligible probability\cite{Cappe2004Dec}. At each time step, the results are weighted using principles from importance sampling to account for the fact that we are sampling from the proposal distribution instead of the prior \cite{Cappe2004Dec}. This algorithm, adapted from \cite{Beaumont2009Nov}, is shown in Algorithm \ref{algo:abc-pmb} and Figure \ref{fig:abc_explained} (the 2 parameter algorithm is equivalent, with the variance replaced by a covariance matrix). The parameters in the algorithm were set following trial and error to balance computation time and accuracy.

We also investigated an alternative approximate Bayesian computation approach known as ABC regression. Instead of the Wasserstein distance, we used the mean of the log transformed event counts as a summary statistic with this method. Full details are in the Supplementary Information.

\begin{figure}[ht]
\centering
\includegraphics[width=.6\linewidth]{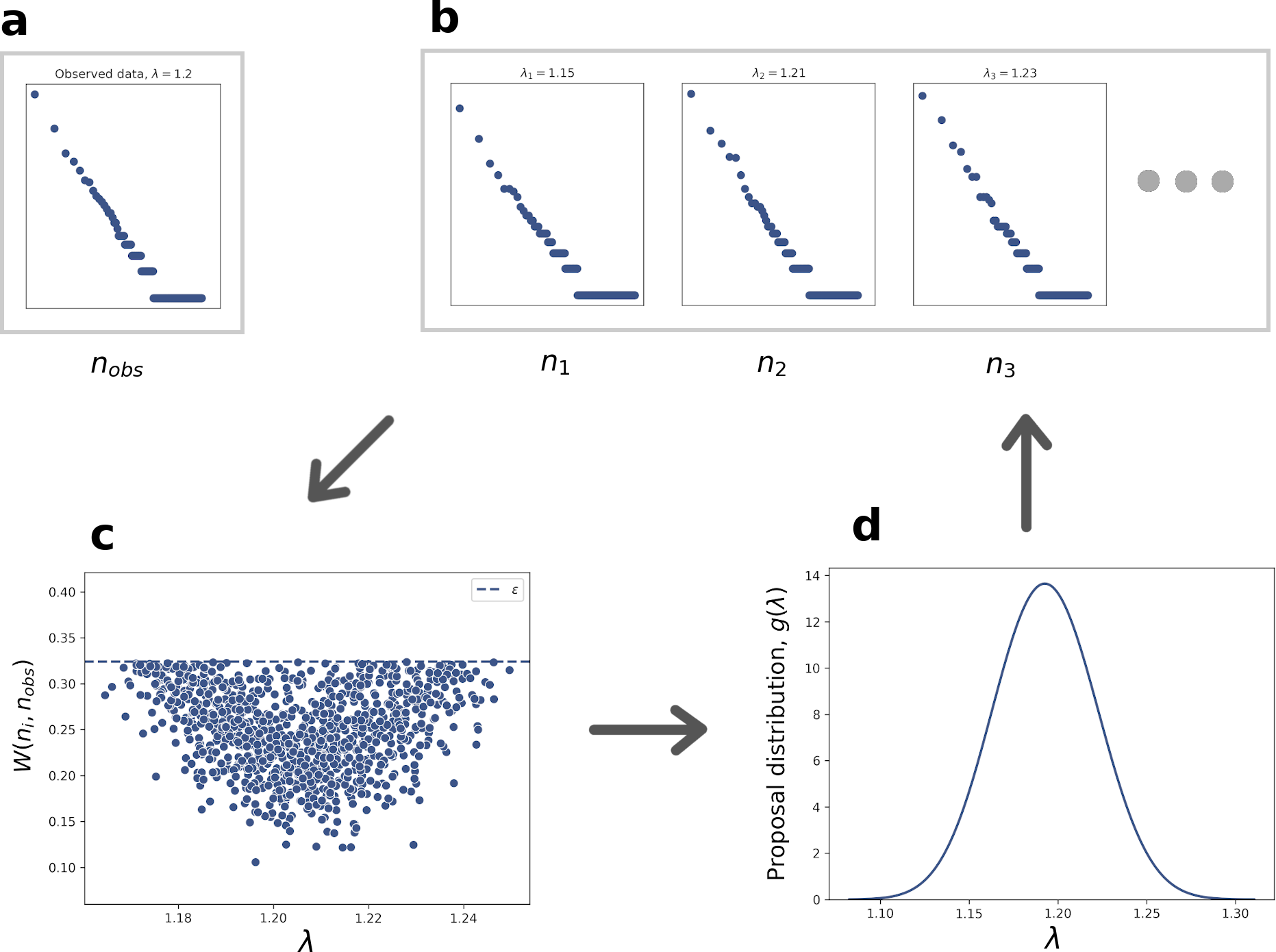}
\caption{ Approximate Bayesian computation with population Monte Carlo (ABC-PMC). a) Given the observed data. b) Particles are generated from a proposal distribution and data is simulated for each particle. For each particle, the Wasserstein distance is measured between the simulated data and the observed data. c) This is repeated until $nParticles$ samples are generated with Wasserstein distance within a tolerance $\epsilon$. d) A new proposal distribution is generated by a weighted kernel density estimate on the accepted particles, with a weighting based on importance sampling principles. A new tolerance is set based upon a proportion of $survivalFraction$ particles with the smallest distances found in this time step. This is repeated for a given number of generations. The final successful particles are used to generate an approximation of the posterior distribution using a weighted kernel density estimate. Figure adapted in part from \cite{Sunnaker2013Jan} and \cite{Csillery2010Jul}.}
\label{fig:abc_explained}
\end{figure}

\begin{algorithm}
\DontPrintSemicolon 
\KwIn{The observed data $\bm{n} = [n_1, n_2, \ldots, n_W], \theta_{min} \gets 1.001, \theta_{max} \gets 3, survivalFraction \gets 0.4, nParticles \gets 256, nGenerations \gets 10$}
\KwOut{Maximum likelihood estimator $\hat{\theta}$}
$priorDist \gets uniformDist(\theta_{min}, \theta_{max})$\;
$nData \gets sum(\bm{n})$\;
$tolerance \gets \infty$\;
$proposalDist \gets priorDist$\;

\For{$g \gets 1$ \textbf{to} $nGenerations$}{
  $\theta s \gets array()$\;
  $ds \gets array()$\;
  $weights \gets array()$\;
  
  \For{$i \gets 1$ \textbf{to} $nParticles$}{  
    $hit \gets FALSE$\;
    \While{$!hit$} {
      $\theta \gets proposalDist.sample()$\;
      \If{$\theta_{min} \leq \theta \leq \theta_{max}$} {

          $z \gets generateData(\theta, nData)$\;
          $d \gets wassersteinDistance(n, z)$\;
          \If{$d \leq tolerance$} {
            $\theta s[i] \gets \theta$\;
            $ds[i] \gets d$\;
            $weights[i] \gets priorDist.evaluate(\theta)/proposalDist.evaluate(\theta)$\;
            $hit \gets TRUE$\;
          }
        }
    }
  }
  $tolerance \gets getTolerance(ds, survivalFraction)$\;
  $var \gets weightedVariance(\theta s, weights)$\;
  $proposalDist \gets KDE(\theta s, weights, bandwidth=sqrt(2 \times var))$\;
}

$posterior \gets KDE(\theta s, weights, bandwidth=sqrt(var))$\;
$\hat{\theta} \gets max(posterior)$\;

\Return{$\hat{\theta}$}\;
\caption{{\sc Approximate Bayesian Computation Population Monte Carlo Zipf's Law}}
\label{algo:abc-pmb}
\end{algorithm}

\section*{ABC Results}

\subsection*{Approximate Bayesian Computation with Zipf Distributions}

Rank-frequency data was generated ($N=$10,000) from an unbounded power law with exponents ranging from 1 to 2. For each generated data set, the exponent was estimated using a) Clauset et al's estimator and b) ABC-PMC with the Wasserstein distance. This was repeated 100 times to find the mean bias and variance. The ABC method has much lower bias and similar variance to Clauset et al's method, (Figure \ref{fig:abc_vs_clauset}).

\begin{figure}[ht]
\centering
\includegraphics[width=.6\linewidth]{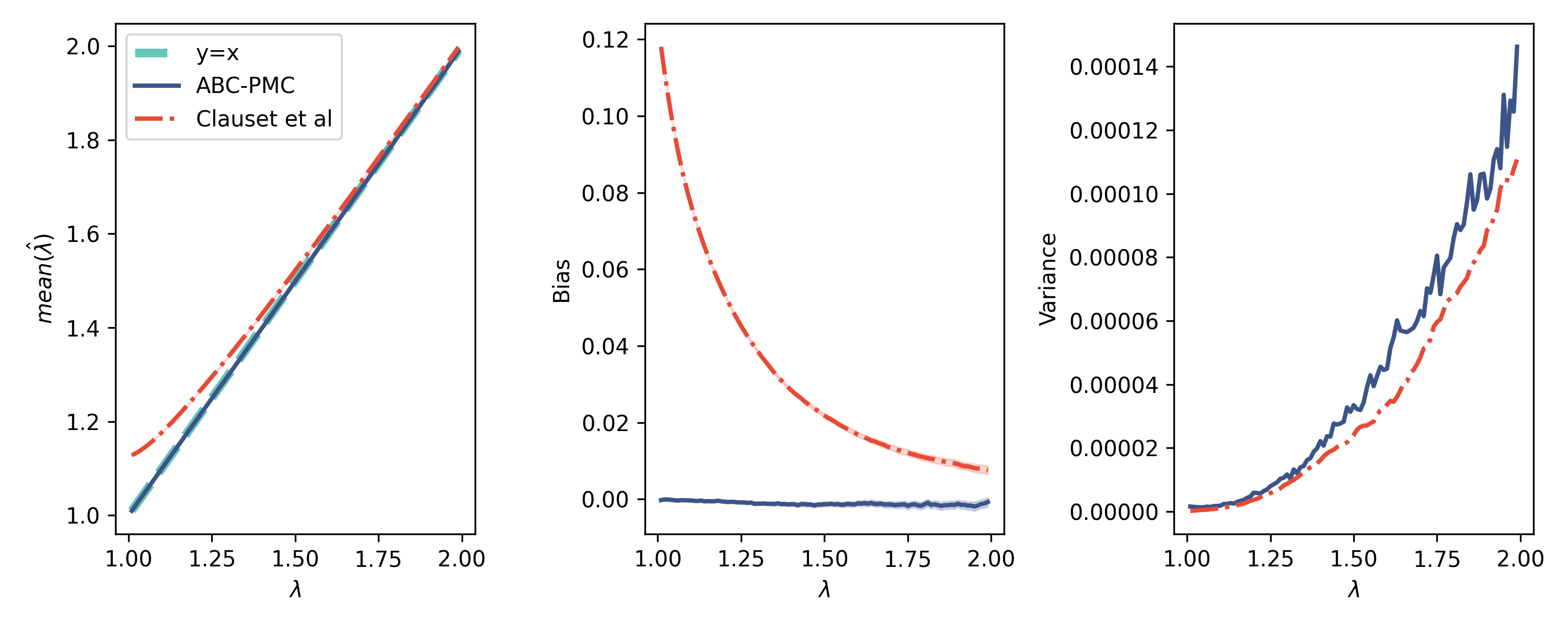}
\caption{ Bias in ABC (solid blue) vs Clauset et al's estimator (dashed red) for unbounded power laws. For each of 100 values of $\lambda$ between 1.01 and 2, rank-frequency data ($N=$10,000) was generated by sampling an unbounded power law. This was run 100 times. The left figure shows the known $\lambda$ and the mean estimated $\lambda$. The centre figure shows the mean bias, with a 68$\%$ confidence interval shaded. The right figure shows the variance of the estimators. The ABC estimator has much lower bias and similar variance to Clauset et al's estimator.} 
\label{fig:abc_vs_clauset}
\end{figure}

We also investigated how the bias changes with varying sample size. Rank-frequency data was generated with $\lambda=1.1$ and varying sample size up to $N=$1,000,000. Clauset et al's estimator shows positive bias at all values of N, although it decreases with large N. ABC shows much lower bias for all values of N. The variance of ABC is higher for $N \lessapprox 1000$. Overall the variance is still very low, and is insignificant compared to the positive bias showed by Clauset et al's estimator (Figure \ref{fig:abc_vs_clauset_by_N}).  

In addition to the results shown here, we explored a variation of the algorithm using ABC rejection with the mean of the logged event counts as a summary statistic. This method has similarly low bias and variance as the results shown here. See the Supplementary Information for full details.

\begin{figure}[ht]
\centering
\includegraphics[width=.6\linewidth]{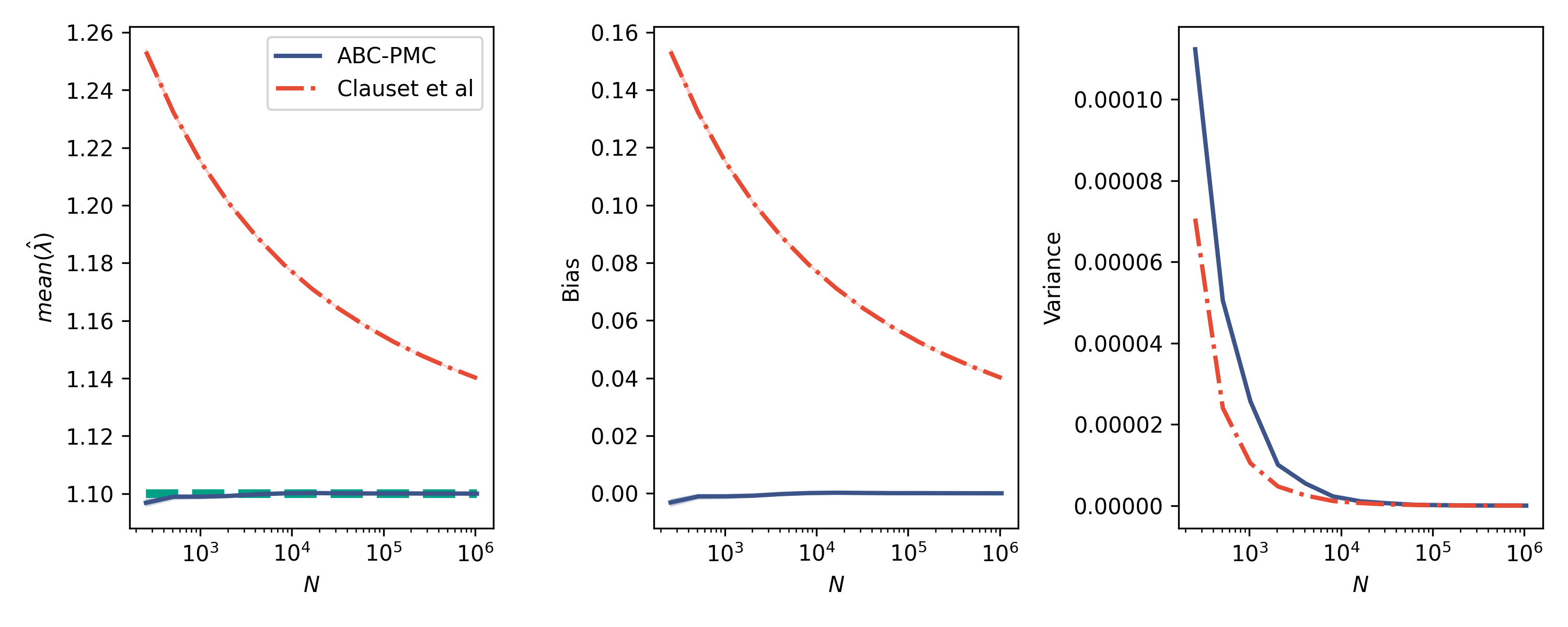}
\caption{ Bias in ABC (solid blue) vs Clauset et al's estimator (dashed red) for unbounded power laws. Rank-frequency data was generated for $\lambda=1.1$ with varying sizes, $N$. This was run 100 times. The left figure shows the known $\lambda$ against the mean estimated $\lambda$. The centre figure shows the mean bias, with a 68\% confidence interval shaded. The right figure shows the variance of the estimators. The bias is much lower with ABC. The ABC estimator has higher variance than Clauset et al at low N, although the variance is still very low. } 
\label{fig:abc_vs_clauset_by_N}
\end{figure}

\subsection*{Approximate Bayesian Computation with Zipf-Mandelbrot Model}

The Zipf-Mandelbrot law is a modification of Zipf's law derived by Mandelbrot that accounts for a departure from a strict power law in the head of the rank-frequency distribution \cite{mandelbrot1953informational},

\begin{equation}
p(r_p) \propto (r_p + q)^{-\lambda} \label{eqn:zipf-manelbrot} \,, \quad q \in [0,1,2 . . . ] \,.
\end{equation} 

\begin{figure}[ht]
\centering
\includegraphics[width=.6\linewidth]{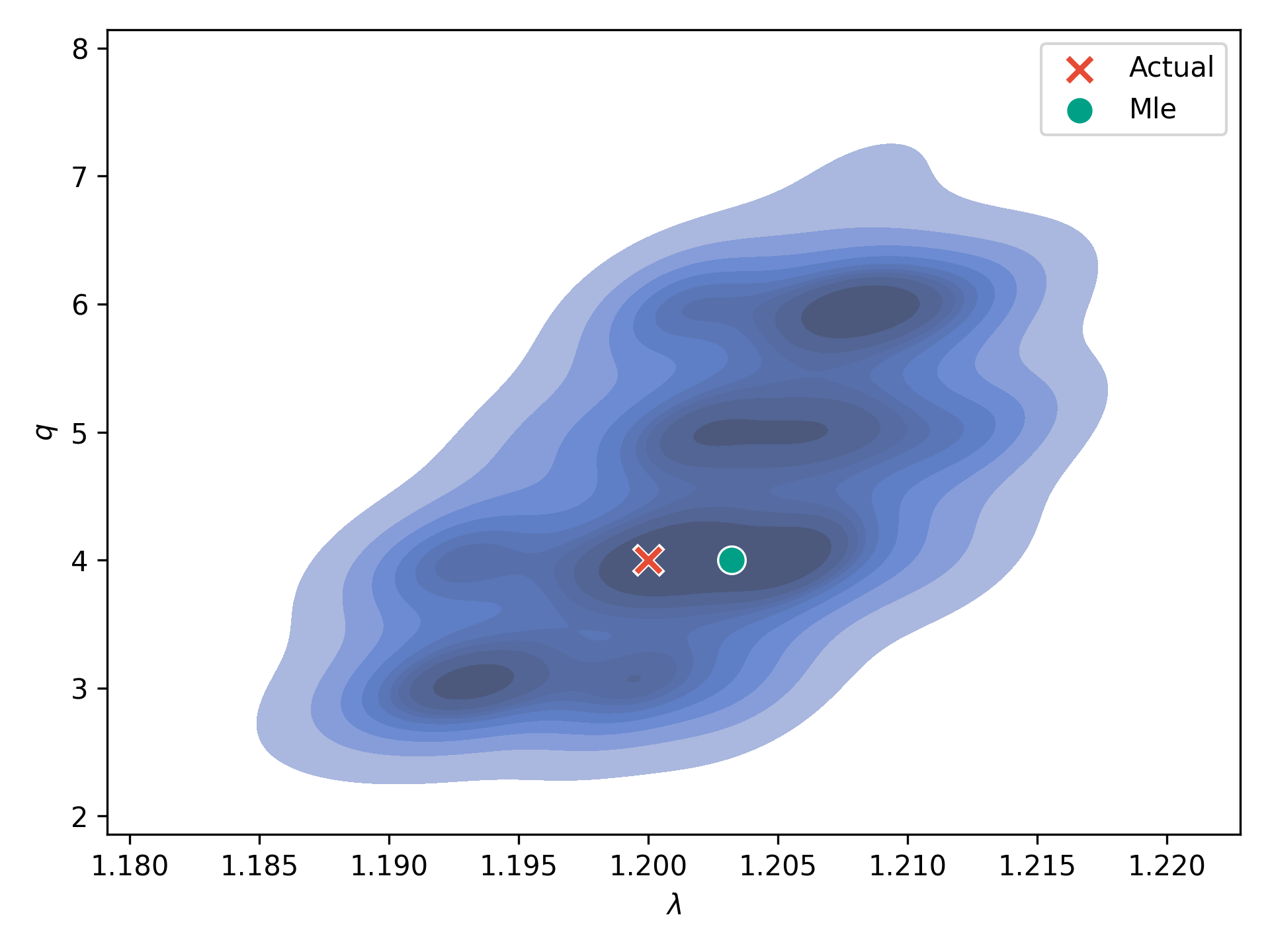}
\caption{Results of ABC-PMC for the Zipf-Mandelbrot law with data generated with known exponent $\lambda=1.2$ and $q=4$ (red cross) with $N=$100,000 words. The likelihood function (darker blue regions have higher likelihood) was approximated using a kernel density estimate. The mode of the KDE gives the maximum likelihood estimate (green circle). The estimator correctly identifies $q$ and is close to the correct exponent $\lambda$.}
\label{fig:zipf_mandelbrot}
\end{figure}

We tested the ABC PMC algorithm with this 2 parameter model. The algorithm is of the same form as Algorithm \ref{algo:abc-pmb}, with the variance replaced with a covariance matrix. The algorithm is demonstrated with one generated data set with $q=$4, $\lambda$=1.2 and $N=$100,000. ABC PMC performs well, with close estimates to the true parameters (see Figure \ref{fig:zipf_mandelbrot}). The approximated likelihood function gives negligible probability for $q=$0, suggesting that the algorithm can discriminate between data generated from Zipf's law and the Zipf-Mandelbrot law. 

\subsection*{Analysis of Books}

Both Clauset et al's method and the approximate Bayesian computation method described here assume a Zipfian data generating model. We have demonstrated that ABC-PMC with the Wasserstein distance works well for data generated from a known power law, with much lower bias than Clasuet et al's method. In the Supplementary Information, we also describe an ABC regression method using the mean log of the word counts that has similar low bias when applied to data from a power law distribution.

It is reasonable to suggest that natural language is a more complex process than drawing words from a power law probability distribution. Indeed, deep learning language models like GPT-3 use billions of parameters \cite{brown2020language}. As such, models that assume Zipfian data generating models are not necessarily suitable for analysing language. To demonstrate the problem, we analysed books using a) Clauset et al's method, b) ABC-PMC with the Wasserstein distance c) ABC regression with the mean of the log transformed word counts as a summary statistic (Table \ref{table:books}). All of the books were downloaded from Project Gutenberg \cite{Gutenberg}. Each text sample was first "cleaned" by removing all punctuation, replacing numbers with a $\#$ symbol, and converting all text to lowercase. The word frequencies were then counted. 

The two forms of ABC give different results, which bracket the results of the Clauset et al estimator. This does not imply that the Clauset et al is the best approximator as we show above that it is biased upwards. What these results indicate is that there is no correct "ground truth" because the assumed underlying models are wrong. 

\begin{table}[!ht]
\centering
\begin{tabular}{|l|l|l|l|}
\hline
\textbf{Book} & \textbf{Clauset et al} & \textbf{ABC PMC with Wasserstein} & \textbf{ABC regression with mean log}\\ \hline
Moby Dick            & 1.19           & 1.25 & 1.16 \\ 
A Tale of Two Cities & 1.21          & 1.27 & 1.17  \\
Alice In Wonderland  & 1.22           & 1.25 & 1.18 \\
Chronicles of London & 1.19           & 1.20 & 1.15 \\
Ulysses              & 1.18           & 1.22 & 1.14\\ \hline
\end{tabular}
\caption{Comparision of estimators of Zipf's law in books.}
\label{table:books}
\end{table}

\section*{Discussion}

We have demonstrated that the prevailing Zipf's law maximum likelihood estimators for rank-frequency data are biased due to an inappropriate likelihood function. This bias is particularly strong in the range of natural language, with exponents close to 1. The correct likelihood function is intractable. We have presented one approach to overcoming this bias using a likelihood-free method of approximate Bayesian computation. The ABC method is shown to work well with data generated from actual power law distributions, with lower bias than Clasuet et al's estimator. 

ABC works well in an idealised situation where the true model is known. However when applied to analysing books, the two ABC approaches that we explored give very different estimates for the Zipf exponents. The Zipfian approaches we investigate all assume a simple bag of words probability model, whereas our results on books indicate that natural language generation is a more complex process--otherwise the two ABC methods would converge. The ABC algorithms are searching a parameter space for the closest model based on the distance measure. This works well when the parameter space includes the true data generating process. But with natural language the assumed simple Zipf model is wrong so there is no "correct" location in the parameter space (or the "correct" location is outside the parameter space). Different distance measures will prejudice different aspects of the observed data and so arrive at different estimates. This bias is arbitrary in nature and there seems to be no reasonable way to decide which distance measure is "correct". The error lies in the assumption of an incorrect data generating model. This problem applies to ABC and Clauset et al's estimator, and seems to be inherent in applying maximum likelihood estimation using simple models to describe rank-frequency power laws in natural language. 

Zipf's law for word types \cite{Corral2019Aug} is an empirical relationship between frequencies of words and ranks in that frequency distribution. The difficulty arises when a probabilistic model is used to describe the mechanism that is generating this relationship, when the actual mechanism is more complex. The main aim of this publication is to clearly show that Clauset et al's estimator is biased for rank-frequency data. The correct likelihood function provides an unbiased framework that works well when the underlying data generating process is known. This does not appear to be the case for natural language. All Zipf estimators have some bias and the best choice will depend on the specific application. Graphical methods such as ordinary least squares may be more suitable to study Zipf's law when investigating the empirical relationship between ranks and frequencies (Equation \ref{eqn:frequency_dist}) and not the probability distribution  (Equation \ref{eqn:prob_dist}). The bias in rank-frequency estimation provides some support for focusing on the alternative frequency-size representation of word counts and Zipf's law for sizes \cite{Corral2019Aug} when studying natural language. 

The scripts and data used here are available at the repository \url{https://github.com/chasmani/PUBLIC_bias_in_zipfs_law_estimators}. That repository includes the approximate Bayesian computation algorithm as well as implementations of other estimators from the literature.

\section*{Acknowledgements}

The study was funded by the EPSRC grant for the Mathematics for Real-World Systems CDT at Warwick (grant number EP/L015374/1).  T.T.H. was supported on this work by the Royal Society Wolfson Research Merit Award (WM160074) and a Fellowship from the Alan Turing Institute, which is funded by EPSRC (grant number EP/N510129/1).

\section*{Author contributions statement}

C.P. conceived of the presented idea and carried out the analyses. T.T.H. supervised C.P. and offered guidance, suggestions and support throughout. All authors reviewed the manuscript. 

\section*{Additional information}

The authors declare no competing interests. 

\bibliography{references}

\section*{Supplementary Information}

\section*{Computational Complexity}

The general likelihood for inferring probability distributions from rank-frequency data is given in the main paper as

\begin{equation}
p(\bm{n}| \bm{\theta}) =  \sum_{\bm{s} \in S(W)} \prod_{r_p}^{W} p(x_{r_p})^{\bm{n}(\bm{s}(r_p))} \label{eqn:likelihood} \,.
\end{equation}

The number of terms in the likelihood function scales as $O(W!)$, so that naive computation of the likelihood is impractical even at $W \approx 10$ . When analysing Zipf's law for words in a book $W$ represents the writer's vocabulary. Even considering a lower bound for $W$ as the number of unique words in a book, $W>1000$ so that the likelihood is extremely computationally expensive using a naive algorithm. Here we will explore how to make this computation more efficient. 

The full likelihood function (equation \ref{eqn:likelihood}) is equivalent to the calculation of the permanent of a matrix with entries $a_{ij}=p(x_j)^{\bm{n}(i)}$: 

\begin{equation}
\centering
A = 
\begin{bmatrix}
p_1^{\bm{n}(1)} & p_2^{\bm{n}(1)} & \dots & p_W^{\bm{n}(1)}\\
p_1^{\bm{n}(2)} & p_2^{\bm{n}(2)} & \dots & p_W^{\bm{n}(2)}\\
\vdots & \vdots & \ddots & \vdots\\
p_1^{\bm{n}(W)} & p_2^{\bm{n}(W)} & \dots & p_W^{\bm{n}(W)}
\end{bmatrix} \,,
\end{equation}

\begin{equation}
\mathcal{L}(\bm{\theta}|\bm{n}, M)) = per(A) \,.
\end{equation}

The permanent is similar to the determinant, with the difference that the negative signs in the Laplace expansion formula for the determinant are all positive \cite{Agrawal2008Jul}. A well known algorithm for exactly computing the permanent of a matrix is Ryser's algorithm \cite{ryser1963combinatorial, Glynn2010Oct} with complexity $O(W2^W)$. The exact computation of the permanent is thought to be $\#P$-hard \cite{valiant1979complexity, Scott2011Dec}, so that no polynomial algorithm exists if $P \neq NP$. A polynomial time approximation algorithm for the permanent of a non-negative matrix (as our matrix is), was discovered by Jerrum et al \cite{jerrum2004polynomial}, with complexity $O(W^{10})$. These algorithms are improvements on the naive case but are still prohibitively computationally expensive for the use case of a text corpora with a vocabulary of $W>1000$. 

We investigated a method of reducing the computational complexity of Ryser's algorithm (in our case) by several orders of magnitude by considering tied empirical ranks, which are equivalent to repeated columns in the matrix $A$. This can be done but the computation time remains extremely prohibitive. A lower bound to an estimate of the computational complexity using this technique would be $O(F2^F)$, where $F$ is the number of unique empirical counts, as the computation would be at least as complex as computing the permanent of a matrix of the unique columns. This would remain prohibitively computationally expensive for real world data sets. The slim hope that remains is to use the structure and symmetry of the matrix to find some shortcut or a reasonable approximation, we leave this as an open question.

\section*{Approximate Bayesian Computation Regression with Mean Log}

Approximate Bayesian computation is a technique for approximating posterior distributions without having to calculate a likelihood function \cite{Sunnaker2013Jan, Beaumont2002Dec, Csillery2010Jul}. Instead, we simulate data, $\bm{n}_i$, from possible parameters, $\lambda_i$, and observe how close that simulated data is to the empirical data (using a distance measure $\rho(\bm{n}_i, \bm{n}_{obs}))$. By looking at the behaviour of simulated data with close distances, we can approximate the posterior distribution, $p(\lambda | \bm{n}_{obs})$.

In order to use ABC to we need a way to measure the "distance" between two data sets. A common technique is to summarise the data sets with a summary statistic, $S(\bm{n})$, and define the distance as the difference between those, $\rho(\bm{n}_i, \bm{n}_{obs}) = S(\bm{n}_i) - S(\bm{n}_{obs})$ \cite{Beaumont2010Nov, Sunnaker2013Jan}. A good summary statistic will capture a lot of information relevant to the likelihood function so that $p(\lambda|\bm{n})\sim p(\lambda|S(\bm{n}))$ \cite{}. With rank-frequency distributions, the mean of the logs of the observations is of a similar form to the likelihood function derived in the main paper. Through experiment this statistic was found to be a good candidate summary statistic,

\begin{equation}
    S_i = \sum_{r_e=1}^{W} \bm{n}_i(r_e) log(r_e) \label{eqn:mean_log} \,.
\end{equation}

There are several flavours of ABC \cite{Beaumont2010Nov, Csillery2010Jul}. Here we use the regression method \cite{Beaumont2002Dec, Csillery2010Jul, Leuenberger2010Jan}.  We only consider distances within some tolerance, $\epsilon$, of the observed data, i.e. $ |S(\bm{n}_i) - S(\bm{n}_{obs})| < \epsilon$. The regression method has advantages over the rejection method that it is computationally more efficient and does not require careful tuning of the tolerance \cite{Beaumont2002Dec}. The key assumption is a linear approximation within the tolerance region:

\begin{equation}
\lambda_i = \beta S(\bm{n}_i) + \alpha + \phi_i  \label{eqn:abc_reg_i} \,.
\end{equation}

Assuming that $\phi$ has an invariant distribution within this tolerance region, we can find estimates  $\hat{\beta}$ and $\hat{\alpha}$ using ordinary least squares regression. To estimate the posterior we are interested in $p(\lambda|S(\bm{n}_{obs}))$, which can be estimated by translating the data points along the regression line,

\begin{equation}
\lambda^{*}_i = \lambda_i - \hat{\beta}(S(\bm{n}_i) - S(\bm{n}_{obs}))  \label{eqn:abc_translate} \,.
\end{equation}

The frequency histogram of these translated points will be approximately proportional to the likelihood function. The histogram can be smoothed using a kernel density estimate and the mode taken to find the maximum likelihood estimator. 
The process is summarised in Figure \ref{fig:abc_explained}.

\begin{figure}[ht]
\centering
\includegraphics[width=.6\linewidth]{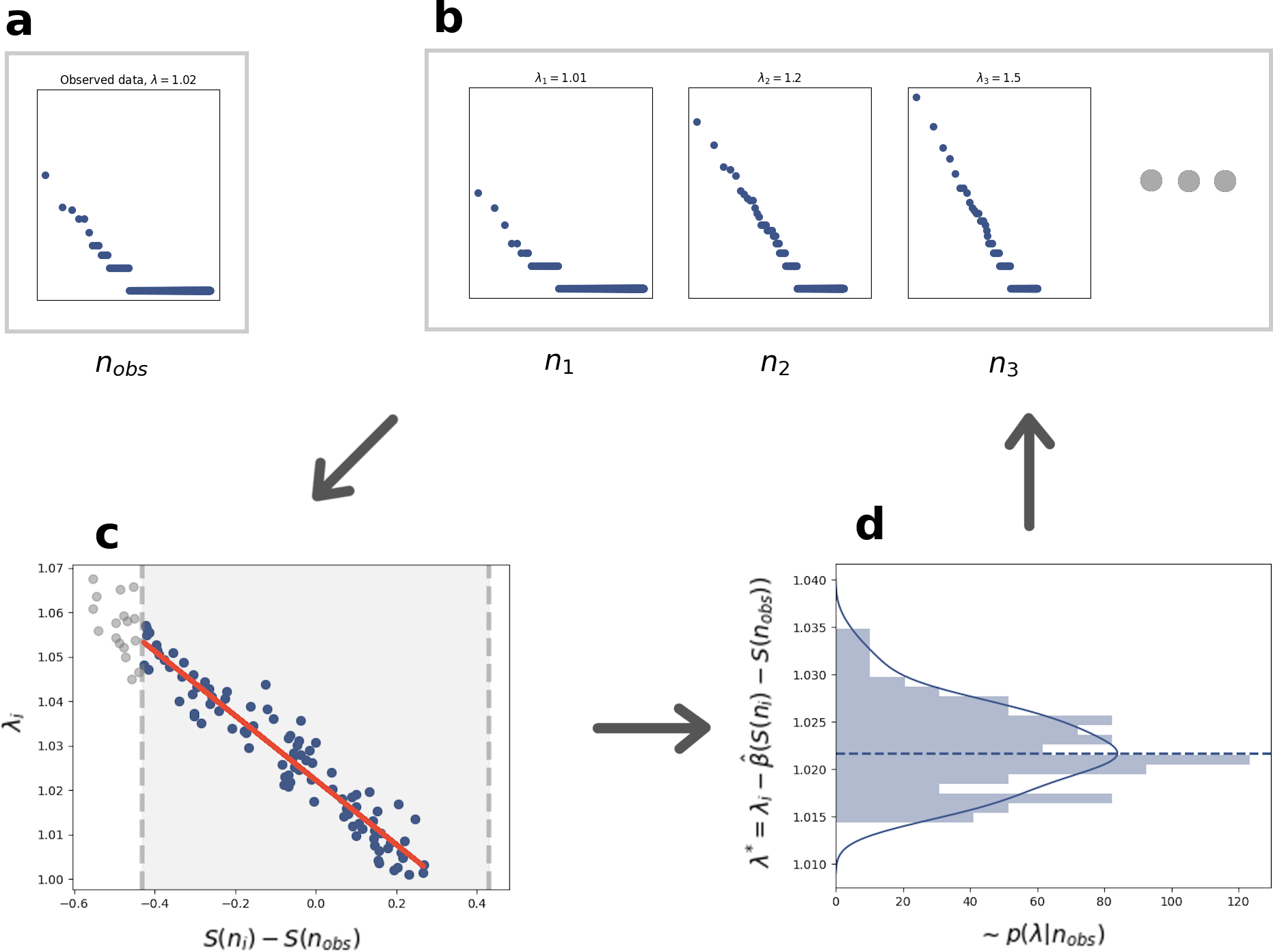}
\caption{ Approximate Bayesian computation regression with the mean log. ABC proceeds as shown. a) A summary statistic $S(\bm{n})$ is calculated from the observed data. b) Parameters are sampled from a uniform distribution. For each parameter, $\lambda_i$ a set of data, $\bm{n}_i$, is generated, and a summary statistic, $S(\bm{n}_i)$, is calculated. c) A tolerance is chosen to accept a given proportion, $P_{\epsilon}$, of the simulations with close summary statistics to the observed data, shown as the shaded region. A linear regression is fit to the accepted simulation results. d) The accepted parameters are adjusted along the regression line to $S(\bm{n}_i) = S(\bm{n_{obs}})$. The histogram of these corrected parameter values approximates the likelihood function. A kernel density estimate is used to smooth the likelihood and find the maximum likelihood estimate for $\lambda$. Here the initial data was generated with $\lambda=1.02$ and the maximum likelihood estimator was $\hat{\lambda} = 1.023$, this is a typical result. Figure idea adapted from \cite{Sunnaker2013Jan} and \cite{Csillery2010Jul}.}
\label{fig:abc_explained}
\end{figure}

\subsection*{ABC Regression Results}

Rank-frequency data was generated ($N=10000$) from an unbounded power law with exponents ranging from 1 to 2. For each generated data set, the exponent was estimated using a) Clauset et al's estimator and b) ABC. This was repeated 100 times to find the mean bias and variance. The ABC method has much lower bias and similar variance to Clauset et al's method, (Figure \ref{fig:abc_vs_clauset}).

\begin{figure}[ht]
\centering
\includegraphics[width=.6\linewidth]{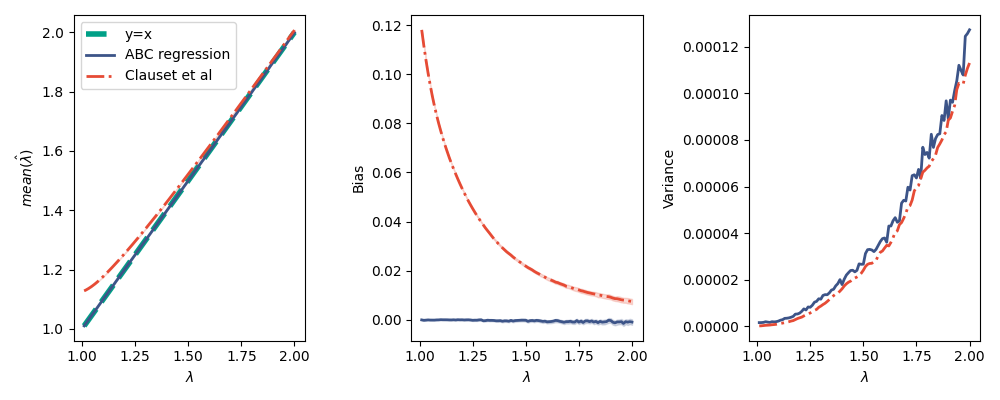}
\caption{ Bias in ABC regression (blue solid line) vs Clauset et al's estimator (red dashed line) for unbounded power laws. Rank-frequency data was generated with $N=10,000$ for 100 values of $\lambda$ between 1.01 and 2. This was run 100 times. The left figure shows the known $\lambda$ against the mean estimated $\hat{\lambda}$ over 100 runs. The central figure shows the mean bias (the difference between the mean estimated $\hat{\lambda}$ and $\lambda$) with a shaded 68\% confidence interval. The right figure shows the variance of the estimators. The ABC estimator has much less bias and similar variance to Clauset et al's estimator.} 
\label{fig:abc_vs_clauset}
\end{figure}

We also looked at changing sample size. Rank-frequency data was generated with $\lambda=1.1$ and varying sample size up to $N=1000000$. Clauset et al's estimator shows positive bias at all values of N, although it decreases with large N. ABC regression shows much less bias at all tested values of N. The variance of ABC regression is higher for $N \lessapprox 1000$. Overall the variance is still very low, and is insignificant compared to the positive bias showed by Clauset et al's estimator (Figure \ref{fig:abc_vs_clauset_by_N}).  

Overall ABC regression with the mean log as a summary statistic shows much less bias and similar variance to Clauset et al's estimator, when applied to data generated from a Zipfian probability distribution.  

\begin{figure}[ht]
\centering
\includegraphics[width=.6\linewidth]{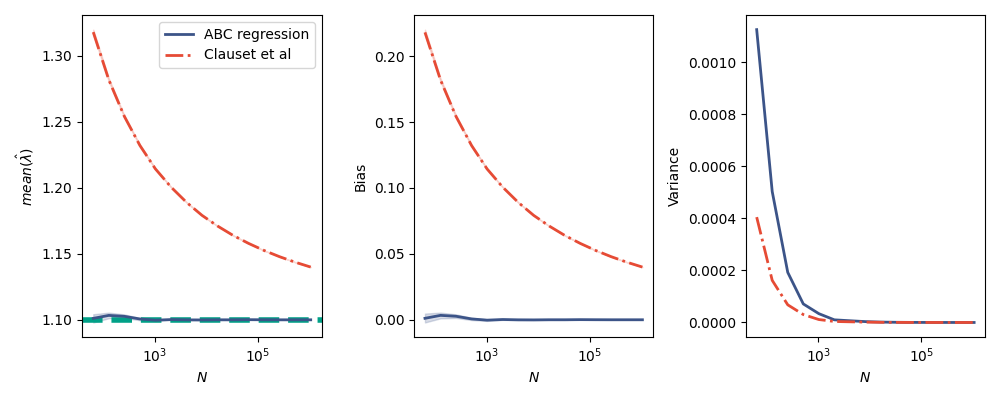}
\caption{ Bias in ABC regression (blue solid line) vs Clauset et al's estimator (red dashed line) for unbounded power laws. Rank-frequency data was generated for $\lambda=1.1$ with varying sizes, $N$. This was run 100 times. The left figure shows the known $\lambda$ against the mean estimated $\hat{\lambda}$. The centre figure shows the mean bias, with a 68\% confidence interval shaded. The right figure shows the variance of the estimators. The ABC estimator has much smaller bias and similar variance to Clauset et al's estimator.} 
\label{fig:abc_vs_clauset_by_N}
\end{figure}

\end{document}